\documentclass{article}
\usepackage{spconf,amsmath,graphicx}

\usepackage{xcolor}
\usepackage{booktabs}
\usepackage{threeparttable}
\usepackage{amsfonts,bm}
\usepackage{mathtools}
\usepackage{pifont}
\usepackage{multirow}

\usepackage{algorithm}
\usepackage{algorithmic}
\usepackage{soul}

\title{IQGAN: Robust Quantum Generative Adversarial Network\\for Image Synthesis On NISQ Devices\vspace{-6pt} \thanks{This work was supported in part by NSF CAREER award CNS-2143120.}}

\name{Cheng Chu\qquad Grant Skipper\qquad  Martin Swany\qquad Fan Chen\vspace{-6pt}}

\address{Luddy School of Informatics, Computing, and Engineering, Indiana University, USA\vspace{-6pt}}

\begin{document}
\maketitle

\begin{abstract}
In this work, we propose IQGAN, a quantum Generative Adversarial Network (GAN) framework for multiqubit image synthesis that can be efficiently implemented on Noisy Intermediate Scale Quantum (NISQ) devices. We investigate the reasons for the inferior generative performance of current quantum GANs in our preliminary study and conclude that an adjustable input encoder is the key to ensuring high-quality data synthesis. We then propose the IQGAN architecture featuring a \textit{trainable multiqubit quantum encoder} that effectively embeds classical data into quantum states. Furthermore, we propose a \textit{compact quantum generator} that significantly reduces the design cost and circuit depth on NISQ devices. Experimental results on both IBM quantum processors and quantum simulators demonstrated that IQGAN outperforms state-of-the-art quantum GANs in qualitative and quantitative evaluation of the generated samples, model convergence, and quantum computing cost. 
\end{abstract}

\begin{keywords}
Quantum machine learning, generative adversarial networks, noisy intermediate-scale quantum
\end{keywords}

\vspace{-8pt}
\section{Introduction}\vspace{-4pt}
\label{sec:intro}
 
\textbf{Related Work and Motivation}.
The recent success of supervised Quantum Neural Networks (QNNs)~\cite{havlivcek2019supervised, huang2022qadv,  tacchino2020quantum, cong2019quantum} has inspired several quantum Generative Adversarial Network (GAN) frameworks~\cite{2013arXivQML, Lloyd2018lett, Pierre2018phyA, Stein2021qugan, niu2021entangling} that extend the application of QNNs to unsupervised generative learning. 
A GAN~\cite{Goodfellow2014GAN, radford2015unsupervised} consists of a generator (G) that generates synthetic samples, and a discriminator (D) that tries to distinguish between true and fake data. Such an adversarial game converges to the point where G generates the same statistics as the true data and D is unable to discriminate between the true and generated data.
Lloyd \textit{et al}.~\cite{Lloyd2018lett} theoretically proved that quantum GANs exhibit potential exponential advantages over classical GANs in high-dimensional data synthesis.
QuGAN18~\cite{Pierre2018phyA} presented an implementation of quantum GAN, which suffers from convergence issues.
QuGAN21~\cite{Stein2021qugan} and EQ-GAN~\cite{niu2021entangling} showed that a quantum fidelity based loss function stabilizes the training process, but 
both works generate data with low quality and require high implementation cost.
In this work, we are motivated to advance previous work~\cite{Stein2021qugan, niu2021entangling} by proposing IQGAN, a robust multiqubit \underline{Q}uantum \underline{G}enerative \underline{A}dversarial \underline{N}etwork architecture for \underline{I}mage synthesis that can be efficiently implemented on today's noisy intermediate-scale quantum (NISQ) devices. We summarize the comparisons between IQGAN and state-of-the-art (SOTA) quantum GANs in Table~\ref{tab:related_work}.

\begin{table}[t!]
\caption{Comparisons between IQGAN and previous works (\textbf{Conv}: Convergence;  \textbf{MultiQ}: multiple-qubit output).}
\label{tab:related_work} 
\footnotesize
\begin{tabular}{|l|l|c|c|l|l|}\hline
\textbf{Scheme}  &\textbf{Loss} &\textbf{Conv}  &\textbf{MultiQ} &\textbf{Quality} &\textbf{Cost} \\\hline\hline
QGAN~\cite{Lloyd2018lett} &N/A &N/A &N/A &N/A &N/A  \\\hline
QuGAN18~\cite{Pierre2018phyA}  &Trace &\text{\ding{56}} &\text{\ding{56}} &Med. &High  \\\hline
QuGAN21~\cite{Stein2021qugan}  &Fidelity &\text{\ding{52}} &\text{\ding{52}} &Low  &High   \\\hline
EQ-GAN~\cite{niu2021entangling} &Fidelity &\text{\ding{52}} &\text{\ding{56}} &Low &Med.	 \\\hline
\textbf{IQGAN} &Fidelity &\text{\ding{52}} &\text{\ding{52}} &High &Low\\\hline
\end{tabular}\vspace{-10pt}
\end{table}


\textbf{Our Contributions}.
This work makes the following contributions.
(1) We study the reasons for the low generative performance in previous work~\cite{Stein2021qugan, niu2021entangling} and conclude that the standard quantum encoders limit the generative ability of a quantum GAN. We then propose a trainable multiqubit quantum encoder that achieves SOTA quality on the generated data (Section~\ref{subsec:encoder});
(2) We present a compact generator circuit ansatz that reduces hardware cost 
and circuit depth 
compared with previous work~\cite{Stein2021qugan, niu2021entangling} (Section~\ref{subsec:generator});
(3) We demonstrate that IQGAN can be efficiently implemented on NISQ devices
and provide the training procedure (Section~\ref{subsec:implement}); 
(4) We evaluate IQGAN on both IBM quantum processors and simulators and show that IQGAN outperforms the prior arts in qualitative and quantitative evaluations of the generated samples, model convergence, and quantum computing cost (Section~\ref{sec:exp}).

\textbf{Limitations and Scope}.
The potential exponential advantage of quantum GANs over classical GANs is proved in~\cite{Lloyd2018lett}. 
In this work, we focus on the situation where the data is classical while the generator and discriminator are quantum, same as~\cite{Stein2021qugan}. We implement a small-scale IQGAN and evaluate its performance against prior arts on MNIST dataset -- a small dataset considered by classical machine learning but SOTA dataset on QNN evaluation~\cite{Stein2021qugan, bausch2020recurrent, wang2022quantumnas, farhi2018classification}. The limitations on the scale of the implementation and the dataset are due to the NISQ technology capabilities. The implementation scale of IQGAN and its corresponding generative performance will improve with the rapidly improving technology. IQGAN demonstrates how NISQ computers can be used to perform nontrivial generative learning tasks with a reasonable performance by overcoming the current quantum obstacles.

\section{Background and Preliminary Analysis}
\label{sec:prelim}
\subsection{Quantum GANs Background}
\begin{figure}[t!]\centering \vspace{0pt}
\includegraphics[width=.88\linewidth]{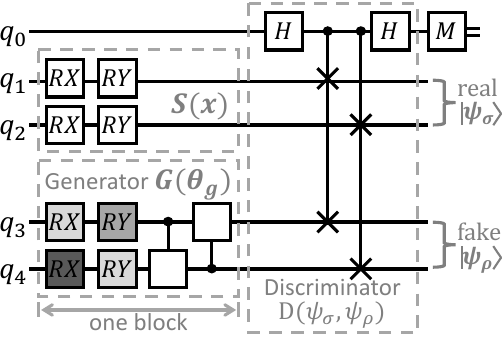}\vspace{-12pt}
\caption{A standard QGAN (\textit{shaded gates are trainable}).}\label{f:background_qgan}\vspace{-16pt}
\end{figure}

\vspace{-4pt}
Figure~\ref{f:background_qgan} shows the mainstream quantum GAN (QGAN) architecture~\cite{Stein2021qugan, niu2021entangling}, which consists of
(1) a data encoder \texttt{S}($\mathbf{x}$) that embeds a real classical input $\mathbf{x}$ to a quantum state $|\psi_\sigma\rangle$;
(2) a Variational Quantum Circuit (VQC) based generator \texttt{G}($\mathbf{\theta_g}$) that generates synthetic data represented as a quantum state $|\psi_\rho(\theta_g)\rangle$;
and (3) a \texttt{SWAP} test based discriminator \texttt{D}($|\psi_\sigma\rangle$, $|\psi_\rho\rangle$) that measures the fidelity between the real data $|\psi_\sigma\rangle$ and the fake data $|\psi_\rho\rangle$.
Due to the limited number of qubits in NISQ devices, high-dimensional inputs are down-sampled using Principal Component Analysis (PCA)~\cite{Stein2021qugan}, while the generated data is obtained by first measuring the qubits of the generator, i.e., $q_3, q_4$ in Figure~\ref{f:background_qgan}, on the probability of the $|0\rangle$ state, and then transforming the measured vector into a high-dimensional data using inverse PCA. 
Note that we consider two-qubit inputs and a \texttt{G}($\mathbf{\theta_g}$) with a single circuit block as an example. The input size and number of blocks in \texttt{G}($\mathbf{\theta_g}$) can be adjusted to fit the problem of interest.

\textbf{Classical-to-Quantum Encoder, \texttt{S}($\mathbf{x}$)}.
Angle encoding is the most widely used method in QNNs~\cite{Stein2021qugan, wang2022quantumnas, farhi2018classification, larose2020robust} due to its noise immunity and simplicity of implementation.
Angle encoding encodes an $N$-dimensional data as the radians of rotation gates acting on $N$ qubits, i.e., 
$|\psi_{\mathbf{x}}\rangle=\bigotimes_{i=0}^{N-1} R(x_i)|0\rangle$,
where \texttt{R} can be one of or a combination of the rotate gates \{\texttt{RX}, \texttt{RY}, \texttt{RZ}\} and $\otimes$ represent the tensor product operation.


\textbf{VQC Circuit Ansatz, \texttt{G}($\mathbf{\theta_g}$)}.
Mainstream QNNs~\cite{wang2022quantumnas, chen2022quantum, qi2022classical} adopt a VQC ansatz constructed by parameterized single-qubit rotation gates followed by nearest-neighbor coupling of qubits using fixed two-qubit \texttt{CNOT} gates, as illustrated in Figure~\ref{f:background_qgan}. 
Such a circuit ansatz has demonstrated superior expressive capability in various applications. 
The logic behind such designs is that single-qubit rotations provide a way to parameterize circuits, while two-qubit \texttt{CNOT} gates provide maximal entanglement between the two target qubits.

\textbf{\texttt{SWAP} Test Circuit, \texttt{D}($|\psi_\sigma\rangle$, $|\psi_\rho\rangle$)}.
A \texttt{SWAP} test circuit is a standard quantum module used for quantum fidelity measurement.
As shown in Figure~\ref{f:background_qgan}, it consists of one ancillary qubit $q_0$, two Hadamard gates (i.e., \texttt{H} gate), and several controlled \texttt{SWAP} gates (i.e., \texttt{CSWAP} gate) that interchange the two quantum states under test only if $q_0$ is in state $|1\rangle$.
The ancillary qubit $q_0$ is finally measured on the z-basis and the probability it yields a measured output $|0\rangle$ is 
$P_0=\frac{1+{\langle{\psi_\sigma}|{\psi_\rho}\rangle}^2}{2}$.
$P_0$ is defined as quantum fidelity as it provides a good estimate of how close two states are, i.e., $P_0=\frac{1}{2}$ for two orthogonal states and $P_0=1$ for two exactly same states.

\textbf{Training Objective}.
The parameter matrix $\mathbf{\theta_g}$ can be equivalently viewed as the weights in classical neural networks and is trained iteratively to minimize a fidelity based loss shown in Equation~\ref{eq:objectFunc}.

\begin{align}\label{eq:objectFunc}\vspace{-24pt}
\begin{split}
\min_{\theta_g} L(\theta_g) = \min_{\theta_g} [1-{\langle{\psi_\sigma}|{\psi_\rho(\theta_g)}\rangle}^2]\\
\end{split}
\end{align}\vspace{-12pt}

\vspace{-10pt}
\subsection{Preliminary Analysis}\vspace{-6pt}
\textbf{Low-quality Generated Images}.
Although prior work~\cite{Stein2021qugan, niu2021entangling} achieved training convergence with the fidelity based loss, their generated data is of low quality.
We show the fidelity between the learned quantum state $|\psi_\rho\rangle$ and the target quantum state $|\psi_\sigma\rangle$ using angle encoding in Equation~\ref{eq:swap_ang_0}$\sim$\ref{eq:swap_ang_1}.
For the sake of simplicity, we highlight a special case of Equation~\ref{eq:swap_ang_1} in Equation~\ref{eq:swap_ang_2} by adding $2n\pi$ to the learned angle $\theta_\rho$.
It is clear that the same fidelity and hence the same loss can be achieved with different parameter sets due to the periodic repeatability of the \texttt{sine} and \texttt{cosine} functions, resulting in distorted generated images (See Figure~\ref{f:dif_work}).
Although input data can be mapped to the range [0,$\pi$] before applying the angle encoding~\cite{Stein2021qugan, chen2022quantum}, such unfaithful normalization prevents a QNN from learning an accurate model and producing high-quality output.
\textit{We conclude that the de facto angle encoding (even with normalization) fails to ensure high-quality output in a fidelity based GAN framework. We investigate if a new quantum encoding method with learnability can improve performance for a quantum GAN.} 

\vspace{-12pt}
{\footnotesize
\begin{align}
\langle\psi_\rho|\psi_\sigma\rangle^2 &= \label{eq:swap_ang_0}
(\begin{bmatrix}
\cos\frac{\theta_\rho}{2} &\sin\frac{\theta_\rho}{2} \\
\end{bmatrix}
\begin{bmatrix}
  \cos\frac{\theta_\sigma}{2}  \\
  \sin\frac{\theta_\sigma}{2}  
\end{bmatrix})^2\\
&= \label{eq:swap_ang_1}
(\cos\frac{\theta_\rho}{2}\cos\frac{\theta_\sigma}{2}+\sin\frac{\theta_\rho}{2}\sin\frac{\theta_\sigma}{2})^2\\
&= \label{eq:swap_ang_2}
(\cos\frac{\theta_\rho+2n\pi}{2}\cos\frac{\theta_\sigma}{2}+\sin\frac{\theta_\rho+2n\pi}{2}\sin\frac{\theta_\sigma}{2})^2
\end{align}
}\vspace{-10pt}

\textbf{Complex and High-Cost Generator}.
The standard circuit ansatz for generator shown in Figure~\ref{f:background_qgan} requires each pair of nearest-neighbors to be maximally entangled using expensive two-qubit \texttt{CNOT} gates.
We hypothesize that 
(1) a flexible amount of entanglement rather than fixed maximal entanglement may perform better for a QNN algorithm;
(2) a generator without entanglement gates may be good enough to provide acceptable generative capability since the encoding circuit \texttt{S}($\mathbf{x}$) normally embeds target data without any entanglement gates.
Therefore, \textit{we proposed to explore the replacement of fixed \texttt{CNOT} gates with other two-qubit gates (e.g., fixed or parameterized), or even completely remove two-qubit gates in a generator circuit. 
We investigate the effect of two-qubit gates on the performance of a generative ansatz and set out to reduce the circuit complexity of a generator.}

\begin{table*}[t]\centering
\caption{Comparison on quantum GANs with different two-qubit gate (2QGate).}\label{tab:results_generator}
\small

\begin{threeparttable}
\begin{tabular}{|c|c|c|c|l|l|l|l|l|l|l|l|}\hline
\textbf{2QGate} &\textbf{Block \#} &\textbf{Normalized Cost} & \textbf{Fidelity} &\textbf{$\theta_0$}  &\textbf{$\theta_1$} 	&\textbf{$\theta_2$}  &\textbf{$\theta_3$}  &\textbf{$\theta_4$}  &\textbf{$\theta_5$} &\textbf{$\theta_6$} &\textbf{$\theta_7$} \\\hline\hline
\texttt{CNOT} &2  &2.8$\times$ &0.813   & \multicolumn{8}{c|}{N/A}  \\\hline
\texttt{ISWAP} &2 &3.7$\times$  &0.954   & \multicolumn{8}{c|}{N/A}  \\\hline
\texttt{CRX}($\theta$)  &2  &6$\times$ &0.969  
&\begin{tabular}[c]{@{}l@{}}-0.007\\ 0.003\end{tabular}
&\begin{tabular}[c]{@{}l@{}}-0.023\\ 0.036 \end{tabular}
&\begin{tabular}[c]{@{}l@{}}-0.228\\ 1.799 \end{tabular}
&\begin{tabular}[c]{@{}l@{}}0.034\\ 0.018 \end{tabular}
&\begin{tabular}[c]{@{}l@{}}-0.093\\ -0.118\end{tabular}
&\begin{tabular}[c]{@{}l@{}}-0.154\\ 0.013 \end{tabular}
&\begin{tabular}[c]{@{}l@{}}-0.015\\ 0.033  \end{tabular}
&\begin{tabular}[c]{@{}l@{}}3.348\\ 3.167  \end{tabular}\\\hline
\texttt{CROT}($\phi$, $\theta$, $\omega$) &2 &6$\times$ &0.969  
&\begin{tabular}[c]{@{}l@{}}0.911\\ 0.079\end{tabular} 
&\begin{tabular}[c]{@{}l@{}}-0.057\\ -0.086\end{tabular} 
&\begin{tabular}[c]{@{}l@{}}2.049\\ -0.068\end{tabular} 
&\begin{tabular}[c]{@{}l@{}}0.059\\ -0.971\end{tabular}
&\begin{tabular}[c]{@{}l@{}}-1.883\\ -0.043\end{tabular}
&\begin{tabular}[c]{@{}l@{}}0.013\\ 0.098\end{tabular}
&\begin{tabular}[c]{@{}l@{}}2.827\\ 0.262\end{tabular}
&\begin{tabular}[c]{@{}l@{}}0.090\\ -0.062\end{tabular}\\\hline
w/o 2QGate &N/A &1$\times$ &0.969  & \multicolumn{8}{c|}{N/A}\\\hline
\end{tabular}
\end{threeparttable}\vspace{-12pt}
\end{table*}

\section{IQGAN: Design and Implementation}\vspace{-8pt}
\label{sec:design}
In this section, we present the design and implementation of the proposed IQGAN framework. The IQGAN architecture has two innovative components:
(1) a variational encoder S($arcsin(\mathbf{x}\cdot\mathbf{\theta_s}$)) that adaptively embeds classical training data into quantum states;
and (2) a compact quantum generator G($\mathbf{\theta_g}$) without costful two-qubit entanglement gates.

\vspace{-8pt}
\subsection{Trainable Multiqubit Quantum Encoder}\vspace{-4pt}
\label{subsec:encoder}
Instead of using a fixed encoder (FE) circuit S($\mathbf{x}$), 
we investigate trainable encoder (TE) using PennyLane~\cite{pennylane} and show that a properly designed variational classical-to-quantum encoder results in improved performance for a QNN model (See results in Section~\ref{sec:exp}). 
We introduce adaptivity into the encoder circuit by defining a variational encoder function  S($arcsin(\mathbf{x}\cdot\mathbf{\theta_s}$)) with trainable parameters $\mathbf{\theta_s}$.
To obtain an angle encoding, we utilize the \texttt{arcsin} function, which is common practice in QNN encoding~\cite{chen2022quantum}.
The parameter set $\theta_s$ is pre-trained to produce faithful quantum presentations in which data from different clusters are separated.
Specifically, the encoder is trained with a pre-train dataset (CIFAR10~\cite{krizhevsky2009learning} in this work) $\mathcal{T}$=\{($\mathbf{x_i}, y_i)|0$$\le$$i$$\le$$N$-1\}, where $\mathbf{x_i}$ is an $n$-dimensional vector and $y_i$ is the label.
We prepare a quantum ensemble $\mathbf{\sigma_{y_k}}$=$\frac{1}{N_k}\sum_{j=0}^{j=N_k-1}|\psi_\sigma(x_j)\rangle\langle\psi_\sigma(x_j)|$ by uniformly sampling $N_k$ inputs from class $y_k$, i.e., $\mathcal{T}_k$=\{($\mathbf{x_j}, y_k)|0$$\le$$j$$\le$$N_k$-1\}, and feeding them into the encoder. 
We then train the encoder to obtain a set of optimal parameters $\theta^*_s$ that maximize the distance between $\mathbf{\sigma_{y_k}}$ and $\mathbf{\sigma_{y_m}}$ when $k{\neq}m$.
IQGAN uses the pre-trained values of S($arcsin(\mathbf{x}\cdot\mathbf{\theta^*_s}$)) as the initial parameters of the encoder.

\vspace{-8pt}
\subsection{Compact Quantum Generator}\vspace{-4pt}
\label{subsec:generator}

To investigate the impact of two-qubit entanglement gates on the performance of a generator circuit ansatz, we replace the default \texttt{CNOT} gates in Figure~\ref{f:background_qgan} with widely used two-qubit gates, including fixed \texttt{ISWAP} gates,
and parameterized \texttt{CRX($\theta$)} and \texttt{CROT($\phi, \theta, \omega$)} gates. 
We refer interested readers to~\cite{gatesdoc} for detailed explanations of each candidate gate. We provide a full comparison between different generator circuit ansatzes in Table~\ref{tab:results_generator}.
The default \texttt{CNOT} gates do not perform very well and actually perform worse than other types of two-qubit entanglement gates.
Parameterized gates, i.e., \texttt{CRX} and \texttt{CROT}, perform better than fixed \texttt{ISWAP} gates (i.e., 0.954 vs. 0.969), which is attributed to the trainable amount of entanglement. However, the hardware overhead of \texttt{CRX} and \texttt{CROT} gates is significantly higher than fixed two-qubit gates.
In particular, we find that a generator without two-qubit gates, denoted as \textit{w/o 2QGate} in Table~\ref{tab:results_generator} achieves the same performance as \texttt{CRX} and \texttt{CROT} with 6$\times$ reduced hardware cost.
Therefore, we propose to implement the generator in IQGAN with no two-qubit gates.

\vspace{-12pt}
\subsection{Implementation of IQGAN}\vspace{-4pt}
\label{subsec:implement}

\noindent
\textbf{Hardware Implementation Cost}.
Multiqubit gates are decomposed in practice into basic single-qubit and two-qubit gates supported by specific gate libraries. We assume an ideal quantum circuit library that supports all types of single-qubit and two-qubit gates and provide a quantitative comparison of design cost between IQGAN and prior arts in Table~\ref{tab:cost}.
IQGAN requires the minimum hardware resource and reduces hardware cost by at least $6\times$ in the simplest case when the generator uses only one VQC block, as detailed in Table~\ref{tab:results_generator}.

\textbf{Training Procedure}.
IQGAN requires a two-step training process. 
At the beginning of the algorithm, the encoder S($arcsin(\mathbf{x}\cdot\mathbf{\theta_s}$)) is initialized by arbitrary parameters $\theta^0_s$ and trained on a pre-train dataset to obtain the optimized parameter set of $\theta^*_s$ as explained in Section~\ref{subsec:encoder}.
We then utilize S($arcsin(\mathbf{x}\cdot\mathbf{\theta_s^*}$)) in an IQGAN framework with uniformly initialized $\theta^0_g$, and train the whole model to minimize the objective function in Equation~\ref{eq:objectFunc}. 
After the model reaches its convergence, we record the optimal parameter set of $\theta^*_g$.

\begin{table}[t]\centering\vspace{-10pt}
\caption{Hardware cost of different quantum GAN schemes ($n$: input size; $b$: repeated VQC block number; 1QG\#/2QG\#: one-/two-qubit gates number; Param.\#: parameter number).}\label{tab:cost}

\small
\begin{tabular}{|l|c|c|c|c|}\hline
\textbf{Scheme} &\textbf{Qubit\#} &\textbf{1QG\#}  &\textbf{2QG\#} &\textbf{Param.\#}  \\\hline\hline
QuGAN21~\cite{Stein2021qugan}	&$2n$+$1$ 	&$nb$+$1$ &$4nb$	&$5nb$ \\\hline
EQ-GAN~\cite{niu2021entangling} &$2n$+$1$ 	&$2nb$+$n$+$2$ &$(b+1)n$	&$2nb$  \\\hline
\textbf{IQGAN} 	&$2n$+$1$ &$2nb$+$n$+$2$ &$n$ &$2nb$	\\\hline
\end{tabular}\vspace{-20pt}
\end{table}


\vspace{-12pt}
\section{Experiments and Results}\vspace{-10pt}
\label{sec:exp}
\begin{figure*}[t]\centering
\includegraphics[width=1\linewidth]{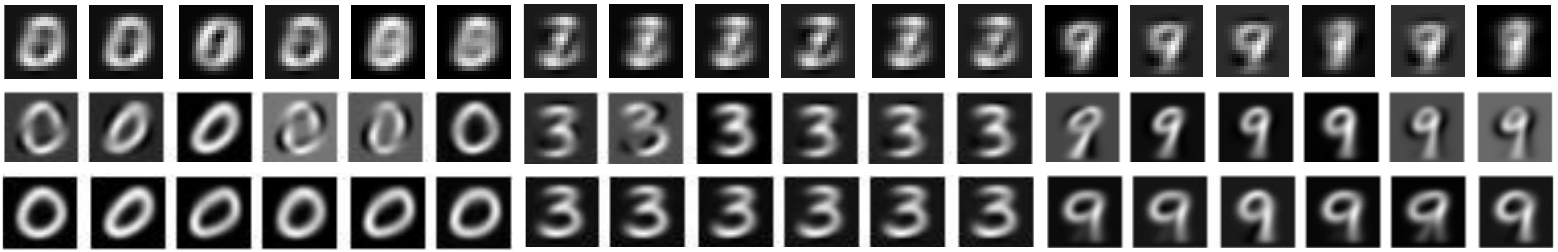}\vspace{-12pt}    
\caption{MNIST images generated by QuGAN2021~\cite{Stein2021qugan} (1st row), EQ-GAN~\cite{niu2021entangling} (2nd row), and IQGAN (bottom row).}\label{f:dif_work}\vspace{-20pt}
\end{figure*}

\subsection{Experimental Setup}\vspace{-6pt}

\textbf{Schemes and Benchmarks}.
We compare IQGAN with two SOTA quantum GANs~\cite{Stein2021qugan, niu2021entangling}
using the MNIST~\cite{mnist} dataset.
Note that we exclude QuGAN2018~\cite{Pierre2018phyA} because it suffers from convergence oscillation and is difficult to train in practice~\cite{niu2021entangling}.
We demonstrate the proof-of-principle of IQGAN using a 5-qubit IBM quantum processor, \texttt{ibmq\_quito}~\cite{ibmq}. 
We confirm that IQGAN is reasonably accurate in generating synthetic images on real NISQ devices. 

\textbf{Simulation}. 
All quantum GANs are implemented using PennyLane~\cite{pennylane}, a quantum computing software library. 
We train QuGAN2021~\cite{Stein2021qugan} using their open-source code by following their training configuration.
We implement EQ-GAN~\cite{niu2021entangling} by following their design strategy but extending their original design to support multiqubit data.
EQ-GAN~\cite{niu2021entangling} and IQGAN are trained with the \texttt{ADAM} optimizer, while the learning rate, batch size, and the number of epochs are set respectively as 0.001, 32, and 30.
The learning rate is scheduled by \texttt{CosineAnnealingLR} with a $T_{max}$ of 30.


\vspace{-12pt}
\subsection{Results and Analysis}\vspace{-4pt}

\begin{table}[t]\centering\vspace{-4pt}
\small
\caption{Accuracy comparison of Fixed Encoder (FE) and Trainable Encoder (TE) on different subsets of MNIST.}\label{tab:results_encoder}
\begin{tabular}{|c|c|c|c|c|}\hline

\multirow{2}{*}{\textbf{Task}} & \textbf{Input} & \multirow{2}{*}{\textbf{Qubit \#}}  & \multicolumn{2}{c|}{\textbf{Accuracy (\%)}} \\\cline{4-5}
&\textbf{Size} &   & \textbf{FE}  & \textbf{TE}  \\\hline\hline
MNIST-2 & 4*4        & 16    & 89.5  & \textbf{90.9} \\\hline
MNIST-4 & 2*2        & 4     & 43.2  & \textbf{45.6} \\\hline
MNIST-4 & 4*4        & 16    & 45.9  & \textbf{49.4} \\\hline
MNIST-8 & 4*4       & 16     & 23.25 & \textbf{24.3} \\\hline
\end{tabular}\vspace{-20pt}
\end{table}

\noindent
\textbf{Effectiveness of the Trainable Encoder}.
To compare the faithful data embedding capability between the \textit{de facto} fixed encoder (FE) and the proposed trainable encoder (TE), we apply both FE and TE to downstream classification tasks and report their accuracy in Table~\ref{tab:results_encoder}.
TE boosts the model accuracy under different input sizes.
We also make two observations that can be extended for general quantum machine learning algorithm:
(1) The effectiveness of TE increases as the input qubit size increases, e.g., the accuracy improvement comparison on MNIST-4 for input sizes of $2\times2$ and $4\times4$;
(2) The effectiveness of TE increases with the model complexity, e.g., the accuracy improvement comparison between MNIST-2 and MNIST-4 with both 16-qubit inputs.

\textbf{Comparison on Image Quality}.
As visualized in Figure~\ref{f:dif_work},
QuGAN2021~\cite{Stein2021qugan} shows limited generative capability. For images of 0 and 9, a blurred outline can be identified. However, for images of 3 with more complex visual structures, QuGAN2021~\cite{Stein2021qugan} fails to reproduce the images.
The results of EQ-GAN~\cite{niu2021entangling} demonstrate an unstable image quality with a large deviation. For instance, the 4th, 5th, 8th, 13th, 17th, and 18th images are highly distorted. 
Note that the results for QuGAN2021~\cite{Stein2021qugan} and EQ-GAN~\cite{niu2021entangling} are simulated data, while the results on IQGAN are collected on \texttt{ibmq\_quito}~\cite{ibmq}.
Compared with previous quantum GANs, we see IQGAN achieves a stable and consistent high-quality output even on NISQ devices.

\begin{figure}[t!]
\begin{minipage}[b]{0.495\linewidth}
\includegraphics[width=1.4in]{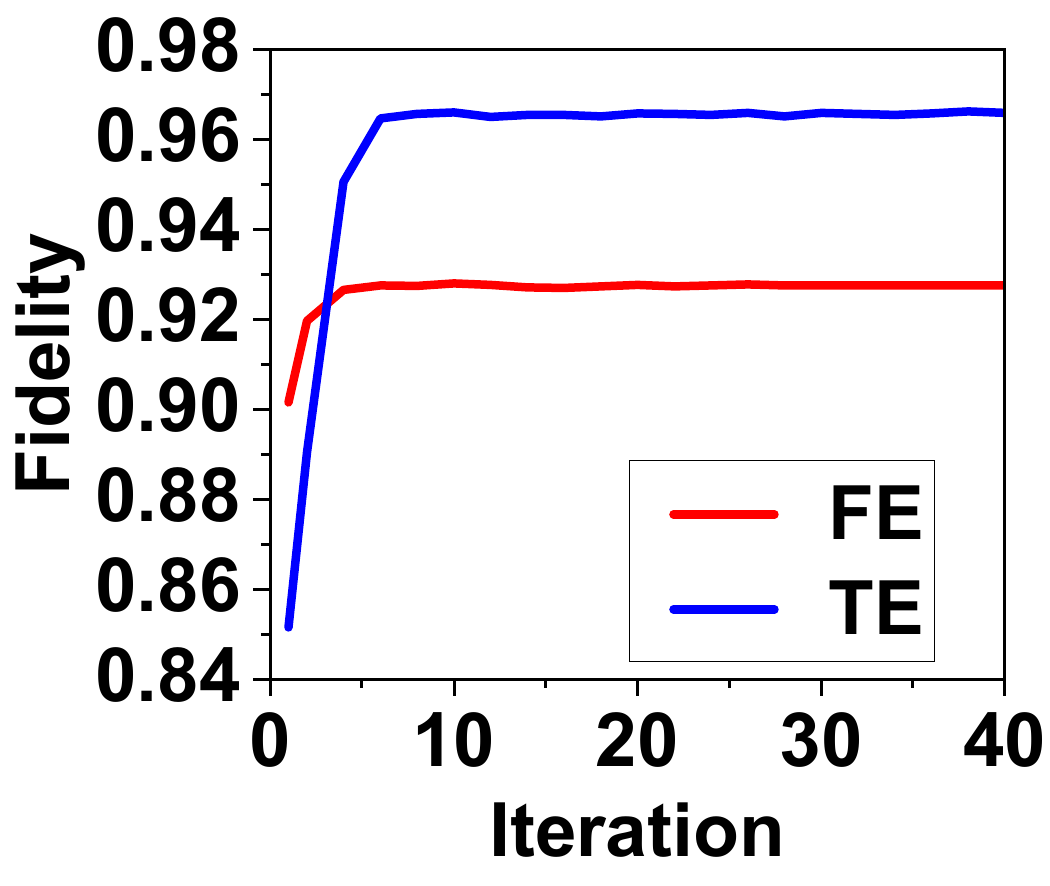}\vspace{-2pt}
\vspace{-0.12in}
\caption{IQGAN Convergence.}\label{f:fidelity_train_encoding}
\end{minipage}\vspace{-8pt}
\hspace{-0.05in}
\begin{minipage}[b]{0.5\linewidth}
\includegraphics[width=1.4in]{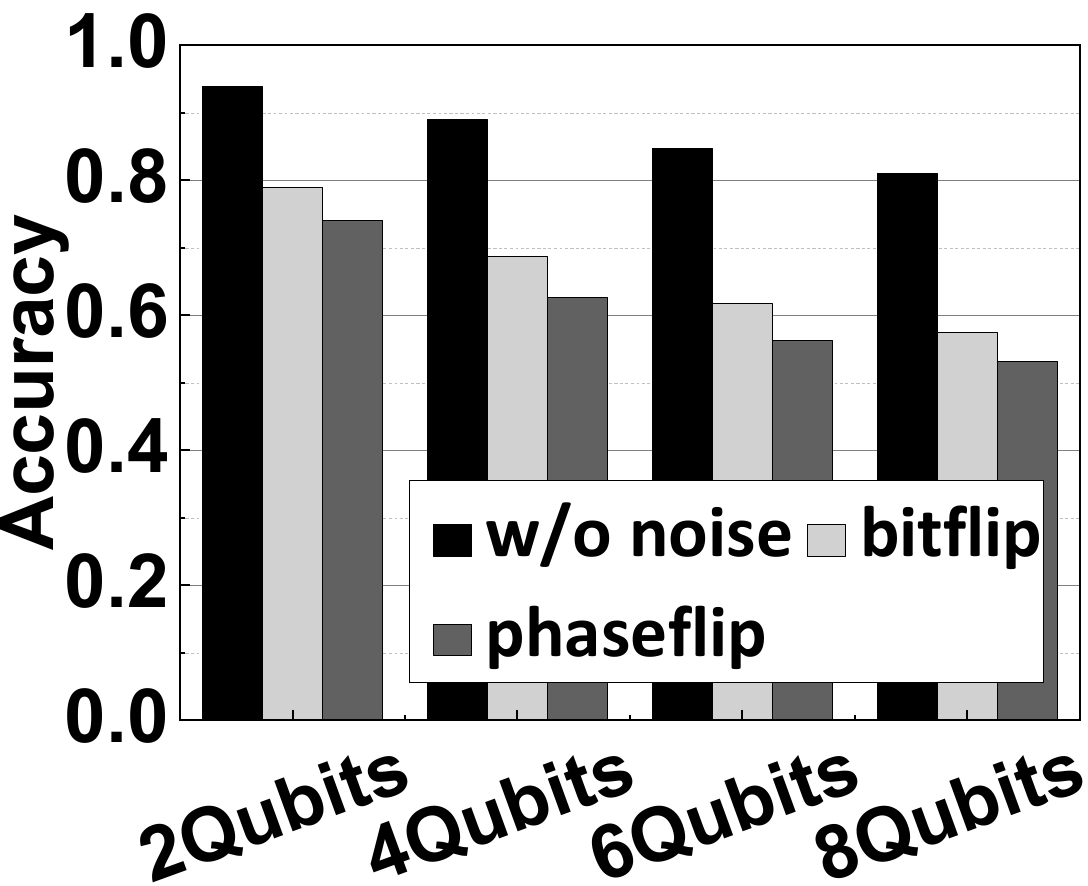}\vspace{-2pt}    
\vspace{-0.12in}
\caption{Impact of noises.}\label{f:dif_qubit}
\end{minipage}\vspace{-12pt}
\end{figure}

\textbf{Convergence of IQGAN}. 
In Figure~\ref{f:fidelity_train_encoding}, we present the measured fidelity of the discriminator in IQGAN. We reported results with both fixed encoder (FE) and trainable encoder (TE).
A stable convergence within ten iterations can be observed for both configurations, confirming that IQGAN converges to a high (i.e., $>$0.92) state overlap with the target inputs.
Compared with a standard FE, the proposed TE achieved a lower (e.g., 0.85 vs. 0.90) fidelity in the first several iterations, however, the final learned fidelity of TE is $\sim$0.039 higher (i.e., 0.966 vs. 0.927) than FE.
Furthermore, the learned fidelity of TE increases faster than that of FE, which is mainly attributed to the higher flexibility and adaptability of the TE circuit and the richer expressive power.

\textbf{Impact of Quantum Errors}.
We also explore the impact of quantum errors on IQGAN performance. We consider bit-flip and phase-flip~\cite{wang2021noise, funcke2020measurement}, which are the two most significant error types on NISQ devices.
We down-sample the original images to 1$\times$2, 1$\times$4, 1$\times$6, and 1$\times$8 vectors using PCA algorithm, and report the fidelity of generated images in Figure~\ref{f:dif_qubit}.
Although the quality of the target images increases monotonically with the input size, the fidelity between learned images and original inputs decreases.
This phenomenon is counterintuitive but can be explained in quantum GAN when quantum errors are considered. 
First, the increased input sizes require an increased number of qubits. 
Second, as the dimension of the data increases, the required generative capability of the generator also increases. Therefore, a more deep generator circuit with an increased number of blocks is required to fit the data.
A quantum system with more qubits and gates is more susceptible to noise.
In this case, while a more complex IQGAN circuit increases the expressive power of the model, quantum noise from excessive overhead reduces the overall fidelity.

\vspace{-12pt}
\section{Conclusion}\vspace{-10pt}
We propose IQGAN, a quantum generative adversarial network for image synthesis that can be implemented on NISQ devices. 
We compared IQGAN with previous work and demonstrate that IQGAN outperforms the state-of-the-arts in image quality and quantum hardware implementation cost.

\newpage
\bibliographystyle{IEEEbib}
\bibliography{main}
\end{document}